\documentclass{article}
\usepackage{graphicx}

\begin{document}

\title{Modeling Melting in Binary Systems}
\author{Leonid D. Son  \\ %EndAName
Ural Pedagogical University, Ekaterinburg 620219, Russia  \and German M.
Rusakov,  \\ %EndAName
Ural Polytechnical Institute, Ekaterinburg 620002, Russia\\ and  \and %
Alexander Z. Patashinski,  \\ %EndAName
Department of Physics and Astronomy and Materials\\ Research Center,
Northwestern University, Evanston IL\\ 60208, USA, and Budker Institute of
Nuclear Physics,\\ Novosibirsk, Russia,\\ and\\ Mark A. Ratner,\\ Department
of Chemistry and Materials Research Center,\\ Northwestern University,
Evanston IL 60208, USA,}
\date{}
\maketitle

\begin{abstract}
A coarsened model for a binary system with limited miscibility of
components is proposed; the system is described in terms of
structural states in small parts of the material. The material is
assumed to have two alternative types of crystalline local
arrangements associated with two components of the alloy.
Fluctuating characteristics of a cluster are the type and the
space orientation of its crystalline arrangement. There are two
different phase transitions in the model system, an orientation
order-disorder transition representing melting, and a phase
transition between phases differing in concentration of
components. Depending on the parameters characterizing the
interaction in the system, this last transition may take place
both in the crystalline and in the amorphous (molten) phase. A
special approximation is used to study the thermodynamics of the
system. The calculated phase diagram describes, at least
qualitatively, the most important features of a binary system.
\end{abstract}
\newpage

\section{Introduction.}

In binary systems, composition-temperature phase diagrams show the regions
of phase stability, usually under constant pressure conditions\cite{PhMet}.
Phase transitions in those systems include melting , which is a universal
phenomenon in all crystalline materials, eventual phase transitions due to
non-miscibility of components and, polymorphous phase transitions that may
appear if there is a competition between different crystalline phases of the
same material. The appearance of these latter transitions depends on the
atomic scale interactions in the material. In isomorphous alloy systems
(e.g. copper -nickel alloy), only a single type of crystal structure is
observed. A more complicated picture appears in alloys with low miscibility
of components. In those systems, clusters having different predominant
components may differ in crystalline arrangements. A known example of an
eutectic system is the copper-silver alloy, while that of a monotectic
system is given by the copper-lead alloy. . Similarities of melting in
different materials allow one to assume that the melting behavior depends on
rather general characteristics (e.g. the particle size, first moments of the
interaction potential) of the interparticle interactions, and that a
coarsened statistical model operating with only a few energy parameters may
reproduce the main features of those materials. In the present paper, we
propose such a coarsened model in terms of structural states of the system
small parts.

In liquids and solids, the interaction of constituent particles is strong
and substantially limits the relative positions of particles. The resulting
strong correlation in small volumes of the material is referred to as the
local order in condensed systems. The term ''local'' refers to a small part
of the material, whose size may be estimated from the correlation functions
as including one or more coordination shells. We refer to this small part as
a cluster. To account for the limitations imposed on relative positions of
cluster particles by the local order, one describes the particle
arrangements in terms of local order parameters. The concept of local order
assumes a coarsening of description to a space scale larger than the
interatomic distance. The resulting model is then a simplification that
allows one to understand the global ordering in terms of the local order.

In simple liquids (e.g. noble gas liquids), radial correlation functions
\cite{HANSEN,Ubbelohde} show some order in the first coordination shell, and
a weaker correlation of particles positions in the second and higher
correlation shells. The possibility to describe these liquids in terms of a
definite local order is an open question. In liquids with stronger
interaction of particles, and especially with covalent or/and hydrogen
bonding, one expects the local order to be stronger, and closer to the local
arrangement in a crystal. This will be referred to as a crystalline local
arrangement.

In the current article, we consider a material in which the local order is
maintained by a strong short range interaction, so that this local order
does not significantly change even when the global order is lost by melting.
In such a material, the loss of the global order during melting results from
proliferation of topological defects, with the density of these defects
small enough to allow recognition of the crystalline local structure. In the
crystalline state, orientations of all clusters almost coincide, while in
the liquid the orientations become uncorrelated above some finite distance.
For a single component material or an isomorphous system with only one type
of local order, a model of melting based on the above physical picture was
studied earlier \cite{MP1}. A more general theory of melting that accounts
for amorphous inherent structures is studied in a recent paper\cite{PR}.

In a crystalline two-component alloy with limited miscibility of components,
one deals at least with two types of local order corresponding to phases
with the predominance of one component. In the materials considered, the
short-range interaction energetically favors, for each sort of atoms, the
same sort of particles in the nearest neighborhood. The physical picture of
the alloy is then a cluster-scale solution. Close to the phase separation
caused by non-miscibility, the material is a solution of one type of
clusters in the matrix of the other type of local structure \cite{PhMet}.
This suggestion determines the choice of cluster states in our model.

In what follows, we propose a model for an alloy based on the above physical
picture. The meta-Hamiltonian of the model is written in terms of a local
order parameter describing two possible predominant configurations in small
elements of the material. The particle arrangement in such a small element
is assumed to be of crystalline type, with orientations of these crystalline
arrangements being the fluctuating characteristics. We study the phase
transitions and the phase diagram in this model.

\section{The local state model of an alloy.}

We divide the material in equal small parts (clusters) of cubic shape. The
centers of these clusters form a simple cubic lattice with sites having
coordinates $r$. In accordance with the discussion in the previous section,
we assume that the relative positions of cluster particles are, within the
accuracy of small thermal displacements and possible packing defects, of a
crystalline type. The crystalline local arrangement is anisotropic. One can
characterize the space orientation of this anisotropy by introducing local
crystalline axes. In the crystal state, the orientations of these axes are
close to the corresponding orientations of some global axes characterizing
the crystal on a macroscopic length scale. In a non-crystalline state, the
orientation of local crystalline axes become, together with the type of the
local crystalline anisotropy and the local composition, a fluctuating degree
of freedom. In our model, we consider only this degrees of freedom. The
universal contribution to thermodynamic properties coming from the small
oscillations of atoms in the vicinity of an energy minimum configuration may
be found, for a classical system, from a rather general consideration\cite
{LANDAU}.

To simplify the modeling, we coarsen the orientation description in a way
proposed in \cite{MP1,PR}. Namely, we divide the orientation space of a
cluster in orientation cells comprising only slightly differing
orientations, and characterize the orientation state of a cluster by the
number ~$i,\ i=\overline{1,N}$~ of the orientation cell. The size of an
orientation cell, and thus the number $N$ of orientation states, will be
discussed later. We assign to a cluster situated at the point $r$ \cite{CHE}
an $N$-component vector $\sigma _j(r)$ by the condition that $\sigma _j(r)=1$
when the orientation state of the cluster is $j$, and $\sigma _j(r)=0$
otherwise. The global state of the material is characterized by the
configuration $\{\sigma _j(r)\}$ and described by the field $\sigma _j(r)$
at all points $r$. Then, a statistical theory of the material may be
formulated in terms of the field $\sigma _j(r)$ of the local order parameter.

One writes the probability $w\{\sigma _i(r)\}$ of a configuration $\{\sigma
_i(r)\}$ as $w\sim \exp [-H\{\sigma _i(r)\}/k_BT]$ where $k_B$ is the
Boltzmann constant, and $H\{\sigma _i(r)\}$ is the meta-Hamiltonian of the
system; below, we choose units in which $k_B$=1. The technique of
meta-Hamiltonians (effective Hamiltonian, nonequilibrium thermodynamic
potential) in the coarsened statistical description of condensed systems was
developed in the theory of critical phenomena (see, e.g. \cite
{LANDAU,PATPOKR} ). To obtain $H\{\sigma _i(r)\}$, one has to take averages,
with Gibbs measure, over all degrees of freedom except for those described
by $\sigma _i(r)$. The meta-Hamiltonian $H\{\sigma _i(r)\}$ thus gives the
free energy of the system in a global state characterized by given local
states $\sigma _i(r)$ of each cluster. The the meta-Hamiltonian $H\{\sigma
_i(r)\}$ depends, through coefficients defining the mathematical form of $%
H\{\sigma _i(r)\},$ on temperature T, pressure P, and other thermodynamic
coordinates of the system. One may use these coefficients as new
thermodynamic coordinates instead of P and/or T. To be specific, we consider
a system at a given pressure P; the thermodynamic coordinates of the state
are the temperature T and component concentrations. A change of the pressure
will change, in general, the meta-Hamiltonian.

The form of the meta-Hamiltonian follows from rather general physical
arguments \cite{LANDAU,PATPOKR}. One expects $H$ to include independent
contributions from single clusters (one-cluster part), each such
contribution $\alpha _i$ depending only on the state $i$ of the
corresponding cluster . The interaction of neighboring clusters is described
by terms depending on states of more than one cluster. Here, we only
consider two-cluster interactions. The reason is that the interacting
potentials acting between particles have a range shorter then the cluster
size, so the interaction is strong only for clusters sharing a border. Then,
for a system in homogeneous external conditions, the general form of the
meta-Hamiltonian is
\begin{equation}
\label{h}H\{\sigma \}=-\sum_{r,r^{\prime }}\sigma _i(r)M_{ij}(r-r^{\prime
})\sigma _j(r^{\prime })-\sum_r\sigma _i(r)\alpha _i(r).
\end{equation}
Here and below, the Einstein's summation rule is supposed for repeating
indices. Let us define the field $\omega _i(r)$ as
\begin{equation}
\begin{array}{c}
\label{proba}<\sigma _i(r)>=\omega _i(r)=\frac 1Z\sum_{\{\sigma \}}\sigma
_i(r)\exp (-\frac{H\{\sigma \}}T),\nonumber \\ Z=\sum_{\{\sigma \}}\exp (-%
\frac{H\{\sigma \}}T);
\end{array}
\end{equation}
the partition function $Z(\alpha _i(r),T)$ is a sum over all local states in
all clusters. The quantity $\omega _i(r)$ defined by (\ref{proba}) is the
probability that the cluster centered at point $r$ is in the state $i$ , it
yields the normalization condition
\begin{equation}
\label{sum}\sum_{i=1}^N\omega _i(r)=1.
\end{equation}
The partition function $Z$ is the generating functional for the field $%
\omega _i(r)$:
\begin{equation}
\label{Zw}\omega _i(r)=T\frac{\delta \ln Z}{\delta \alpha _i(r)}.
\end{equation}

To analyze the thermodynamic behavior of the system, one can make a change
of variables in the functional integral defining the partition function $Z.$
The original meta-Hamiltonian $H\{\sigma _i(r)\}$ depends on a discrete
variable $\sigma _i(r).$ We use the Hubbard - Stratonovich transformation to
re-express the partition function in terms of a continuous lattice variable $%
\psi $. Let us suppose that the matrix $M_{ij}(r)$ has the form:
\begin{equation}
\label{rr}M_{ij}(r)=E_{ij}J(r).
\end{equation}
where $E_{ij}$ is a matrix in the local state space, and $J(r)$ describes
the dependence of interaction on distance between clusters. Introducing a
real, $N$-component ''conjugated'' field $\psi _i(r)$, one rewrites the
expression for $Z$ as
\begin{equation}
\begin{array}{c}
Z=\int \prod_{i;r}D\psi _i(r)\exp \{-\frac 1{2T}\sum_{r,r^{\prime
}}J^{-1}(r-r^{\prime })E_{ij}\psi _i(r)\psi _j(r^{\prime })\}\times
\nonumber \\ \times \prod_{i;r}Tr\exp [\frac 1T(E_{ij}\psi _i(r)\sigma
_j(r)+\alpha _j(r)\sigma _j(r))].\label{Z1}
\end{array}
\end{equation}
Here, the functional integration $D\psi $ goes over all components $\psi _i$
of the field $\psi _i(r)$ . The integration path in the complex $\psi $%
-space may, for a non- positive definite matrix $M_{ij}(r),$ deviate from
the real axes to make the integral convergent. The advantage of the form (%
\ref{Z1}) is that now the summation over local states can be easily taken:
\begin{equation}
\begin{array}{c}
Z=\int \prod_{i;r}D\psi _i(r)\exp \{-\frac FT\},
\nonumber \\ F=\frac 12\sum_{r,r^{\prime }}J^{-1}(r-r^{\prime })E_{ij}\psi
_i(r)\psi _j(r^{\prime })-
\nonumber \\ -T\sum_r\ln [\sum_i\exp \{\frac 1T(E_{ij}\psi _j(r)+\alpha
_i(r))\}].\label{Z2}
\end{array}
\end{equation}
The new variable $\psi _i(r)$ is a continuous one, so one can study the
above expression in the saddle-point approximation, in which the only
allowed configuration is the saddle point one. In the next approximation one
studies then the fluctuations in the vicinity of the saddle point
configuration, to find the first fluctuation corrections to the saddle-point
approximation. The saddle-point approximation is known to fail at critical
points; for phase transitions of the first Ehrenfest order, the saddle-point
approximation may be expected to adequately describe the thermodynamics of
the system and phase transitions. In the present article we shall restrict
ourselves with the first saddle-point approximation.

The saddle-point configuration $\psi (r)$ minimizes the thermodynamic
potential $F$ in (\ref{Z2}), it obeys the equation:
\begin{equation}
\begin{array}{c}
\sum_{r^{\prime }}J^{-1}(r-r^{\prime })\psi _i(r^{\prime })=Z_0^{-1}\exp
[\frac 1T(E_{ij}\psi _j(r)+\alpha _i)],
\nonumber \\ Z_0=\sum_i\exp [\frac 1T(E_{ij}\psi _j(r)+\alpha _i)].\label
{eq1}
\end{array}
\end{equation}
To obtain the generating functional $Z$ one substitutes the
solution of this equation in (\ref{Z2}). For the probability
$\omega_i(r)=-\delta F/\delta
\alpha _i(r)$ one has
\begin{equation}
\begin{array}{c}
\omega _i(r)=<\sigma _i(r)>=\sum_{r^{\prime }}J^{-1}(r-r^{\prime })<\psi
_i(r^{\prime })>.\label{w3}
\end{array}
\end{equation}
In the saddle-point approximation, the quantity $F$ in formula (\ref{Z2})
plays the role of the nonequilibrium thermodynamic potential \cite
{LANDAU,PATPOKR}; at equilibrium, $F$ has a minimum. We define $\omega _i(r)$
for nonequilibrium values of $\psi $ by the formula (\ref{w3}). Then, the
nonequilibrium potential $F$ may be written in terms of $\omega _i(r)$
\begin{equation}
\begin{array}{c}
F=\frac 12\sum_{r,r^{\prime }}M_{ij}(r-r^{\prime })\omega _i(r)\omega
_j(r^{\prime })-
\nonumber \\ -T\sum_r\ln [\sum_i\exp (\frac 1T(\sum_{r^{\prime
}}M_{ij}(r-r^{\prime })\omega _j(r^{\prime })+\alpha _i))].\label{Gamma1}
\end{array}
\end{equation}
The quantity $\omega _i(r)$ may be interpreted as the probability
to find the cluster in the $i$-th state. Because of this
probabilistic character of the quantity $\omega_i(r)$ , we will
call the saddle point approximation the mean-probability
approximation (MPA). The MPA approximation is given by the
equations
\begin{equation}
\begin{array}{c}
\frac{\delta F}{\delta \omega _i(r)}=0,\label{eq2}
\end{array}
\end{equation}
with an additional condition (\ref{sum}). In the explicit form, this
equations may be written as
\begin{equation}
\begin{array}{c}
\omega _i(r)=Z^{-1}(r)\exp (
\frac{E_i(r)}T)\nonumber \\ E_i(r)=\sum_{r^{\prime }}M_{ij}(r-r^{\prime
})\omega _j(r^{\prime })+\alpha _i(r)
\nonumber \\ Z(r)=\sum_{i=1}^N\exp (\frac{E_i(r)}T).\label{MPA}
\end{array}
\end{equation}
Here, $-E_i(r)$ may be interpreted as the energy of the cluster when the
states of surrounding clusters are characterized by the mean probability $%
\omega_i(r^{\prime})$. If the field $\alpha _i$ is spatially homogeneous $%
(\alpha _i(r)=\alpha _i)$, the quantities $E_i,\omega_i$ are, too,
spatially homogeneous:
\begin{equation}
\begin{array}{c}
E_i=M_{ij}\omega_j+\alpha_i,\ \ \
M_{ij}=\sum_rM_{ij}(r).\label{E1}
\end{array}
\end{equation}
The normalization condition (\ref{sum}) holds for solutions of (\ref{MPA})
automatically. The equations (\ref{MPA}) have, in general, more than one
solution; the number of solutions depends on the values of the parameters $%
M_{ij},\alpha _i,T.$ As mentioned above, the characteristics $M_{ij},\alpha
_i$ depend on thermodynamic coordinates of the state (temperature, pressure
etc.), and may be used as new thermodynamic variables. Different solutions
correspond to different phases of the system. In the vicinity of the first
order phase transition, one expects at least two minima of $F$,
corresponding to the stable and metastable phases, with the lowest minima
corresponding to the stable phase. By specifying the form of the matrix $%
M_{ij}$, one arrives at different models. Some of these models are discussed
in following sections.

\section{Melting of a single-component material, and the Potts model.}

In this section, we discuss a coarsened model for melting in a
single-component material \cite{MP1,PR}. The discussion serves as an
introduction to the modeling in the next section; a more detailed study of
one-component melting may be found in ref.\cite{PR}. Here, the material is
supposed to preserve the same crystalline local order both in the crystal
and in the liquid state. The only degrees of freedom described by the model
are the local orientations. To give a coarsened description of orientation
order-disorder, one divides the cluster orientation space in angular cells.
By definition, all orientations within a given angular cell correspond to
the same cluster orientation state, labelled by the angular cell number. The
number of cluster orientation states is then determined by the size of an
angular cell. With increase of the angular cell size, the description
becomes more coarse. We choose this size to model the main features of the
system in a most schematic way.

In accordance with the experimental data, the ground state ( the state
having the lowest energy) is a crystal, with all clusters occupying the same
angular cell. The globally amorphous (but locally crystalline in materials
under consideration) global state of the material has a higher energy. This
implies an angular attraction between neighboring clusters: the
meta-Hamiltonian is at minimum when their orientations coincide and all
clusters form an ideal crystal. One could expect that this angular
attraction has an angular size, so that the two-cluster contribution to the
meta-Hamiltonian increases sharply with the disorientation of the clusters,
and then substantially flattens when the mutual disorientation becomes
larger than the angular attraction size. We choose the size of the angular
cell equal the angular attraction region. Then, a simple schematic way to
model the energetics of neighboring cluster interactions is to assign an
angular binding energy $J<0$ to two neighboring clusters belonging to the
same orientation cell (having the same orientation state). The interaction
energy of two neighboring clusters having different orientation states is
chosen as the origin of the energy scale $E=0.$ The resulting model with
only one angular energy characteristics has the form of an $N$ state Potts
model with the matrix $M_{ij}$ having the form
\begin{equation}
\begin{array}{c}
M_{ij}=\delta _{ij}J(r)~,
\nonumber \\ \label{diag}
\end{array}
\end{equation}
The assumed equivalence of all orientation states leads to a high symmetry .
The parameter $\alpha _i$ is now state-independent and may be put equal to
zero without loss of generality. In the symmetric high-temperature phase,
all $N$ equivalent states have equal probabilities
\begin{equation}
\omega _{i,h}=\frac 1N,\ \ \ i=1,...,N.
\end{equation}
At lower temperatures, this symmetry is spontaneously broken. At
these temperatures, one of the states, which we label by $i=1$, is
more probable than the others ( $\omega_1>1/N$), while N-1 states
$i$=2,... ,N have equal probabilities
\begin{equation}
\omega _i\mid _{i\neq 1}=\frac{1-\omega _1}{N-1}.
\end{equation}

From the minimum condition for the non-equilibrium potential F
(see (\ref {Gamma1})) one obtains the MPA equation (\ref{MPA})
(see
\cite{MP1,PR} )
\begin{equation}
\omega _1=Z^{-1}e^{J\nu \omega _1/T},Z=e^{J\nu \omega _1/T}+(N-1)e^{J\nu
(1-\omega _1)/(N-1)T}.
\end{equation}
Here, $\nu J(0)=\sum_rJ(r),$ with $\nu $ being the effective number of
cluster neighbors. Depending on the temperature, the MPA equation has
different solutions describing the high-temperature and the low temperature
phases. In the high-temperature phase, all states have equal probabilities $%
\omega _{i,h}=1/N,$ so that the quantity $x=\omega _1-\frac 1N$ equals zero.
This solution ($x=0$) formally exists at all temperatures but corresponds to
a minimum of ~$F$, and thus to a stable or metastable state,~only at ~$T>T_2$%
. The low - temperature solution ~$x\approx (N-1)/N$~ exists at ~$T<T_1$.
The transition from the low-temperature ($x\neq 0$) to the high temperature,
symmetric phase ($x=0$) takes place at ~$T=T_m,$ $T_1>T_m>T_2$; this
transition is of the first Ehrenfest order (discontinuous). For ~$N\gg 1$~
the temperatures $T_1,T_2,T_m$ may be found in explicit form:
\begin{equation}
\label{TEMP}T_1\approx \frac{\nu J(0)}{\ln N},T_2\approx \frac{\nu J(0)}%
N,T_m\approx \frac{\nu J(0)(N-1)}{2N\ln N}.
\end{equation}
The model may be applied to chemical compositions of stoichiometric
compounds behaving like pure materials \cite{Westbrook}. The necessary
condition of applicability is that the local order should not be destroyed
via melting, including the chemical order (the material should not
dissociate).

In multicomponent materials there may be a competition of several types of
local structure. In a binary system ~$A-B$~, these competing structures may
correspond to pure components and to all intermediate stoichiometric
compositions. The existence of competing structures enhances the set of
local states. In the following section , we generalize the above model to
account for two competing types of local order.

\section{System with two competing local structures.}

In this section, we propose and discuss an idealized model of a material
with two possible local structures. To account for the type of local
structure as a new degree of freedom, we introduce two indices for labelling
the local state. The vector $\sigma $ is now written as
\begin{equation}
\sigma _i^j;\ \ \ j=1,2.
\end{equation}
The upper index corresponds to the type of local order, while the lower one
numerates the orientations of a cluster. As in the previous section, we
assume division of the orientation space into orientation cells, with $n$
orientation states for the first and $m$ for the second local structure. The
meta-Hamiltonian of the model may be written in the form
\begin{equation}
\label{H}H=-\alpha \sum_r\sum_{i=1}^n\sigma _i^1(r)-\sum_{r,r^{\prime
}}\sigma _i^k(r)M_{il}^{kj}(r-r^{^{\prime }})\sigma _l^j(r^{^{\prime }}).
\end{equation}
The parameter $\alpha $ describes the internal energy difference between two
local structures. As in the case of a one-component system, we choose the
angular cell sizes to comprise the angular attraction regions. The kernel $%
M_{il}^{kj}(r)$ of interaction is assumed to differ from zero only for
nearest neighbors, and to have the form
\begin{equation}
\label{M}M_{il}^{11}(r)=\tilde J_1(r)\delta _{il};\ M_{il}^{22}(r)=\tilde
J_2(r)\delta _{il};\ \ M_{il}^{12}(r)=M_{il}^{21}(r)=\tilde \varepsilon (r)
\end{equation}
The interaction between clusters of same composition is chosen in the Potts
form discussed in the previous section. The interaction between clusters
having different local structures (the term $M^{12}$ and $M^{21}$) is
assumed to be orientation-independent; the argument here is that if clusters
with different structures share a boundary, then the energy of the boundary $%
\tilde \varepsilon $~ is mostly due to the difference in the neighboring
structures and less due to their relative orientations. Parameters ~$\tilde
J_1,\tilde J_2$~ characterize then the orientational interaction of same
type clusters, and ~$\tilde \varepsilon $ is an inter-structural ''surface''
energy. We will denote
\begin{equation}
\varepsilon =\int \tilde \varepsilon (r)dV;\ \ \ J_1=\int J_1(r)dV;\ \ \
J_2=\int \tilde J_2(r)dV.
\end{equation}

>From symmetry arguments, one has for the mean values of the parameter $%
\omega $ (see the previous section):
\begin{equation}
\begin{array}{c}
\omega _1^1=\omega _1;\ \ \ \omega _{k\neq 1}^1=
\frac{x-\omega _1}{n-1} \\ \omega _1^2=\omega _2;\ \ \ \omega _{k\neq 1}^2=%
\frac{1-x-\omega _2}{m-1}.\label{sigma}
\end{array}
\end{equation}
Here, ~$x$~ is the concentration of clusters having the first
structure. The quantities $\omega_1$and $\omega_2$ in
(\ref{sigma}) describe the orientational ordering; the lower index
1 labels the preferred orientation in the orientationally ordered
(crystalline) state. Now one is able to classify the possible
phases in the model:

\begin{enumerate}
\item  $\omega _1\sim x,\ x\sim 1$~ - crystal with the first type of local
structure (C1);

\item  $\omega _1=x/n,\ x\sim 1$~ - liquid with the first type of local
structure (L1);

\item  $\omega _2\sim 1-x,x\sim 0$~ - crystal with the second type of local
structure (C2);

\item  $\omega _2=(1-x)/m,x\sim 0$~ - liquid with second type of local
structure (L2)
\end{enumerate}

The nonequilibrium thermodynamic potential $F$ defined by formula (\ref
{Gamma1}) may be calculated using the meta-Hamiltonian of the model; some
lengthy algebra then leads to the form
\begin{equation}
\begin{array}{c}
F=\frac 12(J_1\omega _1^2+J_1
\frac{(x-\omega _1)^2}{n-1}+J_2\omega _2^2+J_2\frac{(1-x-\omega _2)^2}{m-1}%
)+ \\ +\varepsilon x(1-x)-T\ln \{\exp (
\frac{J_1}T\omega _1+\frac \varepsilon T(1-x)+\frac \alpha T)+ \\ +(n-1)\exp
(
\frac{J_1(x-\omega _1)}{T(n-1)}+\frac \varepsilon T(1-x)+\frac \alpha T)+ \\
+\exp (\frac{J_2}T\omega _2+\frac \varepsilon Tx)+(m-1)\exp (\frac{%
J_2(1-x-\omega _2)}{T(m-1)}+\frac \varepsilon Tx)\}\label{fff}
\end{array}
\end{equation}
The minima of this function determine equilibrium phases and the phase
diagram of the model; we study the phase boundaries in the following section.

\section{Binary system with limited miscibility.}

Consider a binary system in which the two local structures correspond to
phases with predominance of component $A$ and $B$ respectively. If these
phases are stoichiometric compositions, the model describes the material in
some range of intermediate concentrations; we assume that no other
stoichiometric composition can add a third type of local order. We consider
systems in which the two types of local order differ sharply (have different
groups of local symmetry or incompatible interatomic distances). Systems
with limited miscibility usually exhibit a sharp decrease of melting
temperature with deviation from the stoichiometric compound \cite
{Westbrook,Goldsmith}. The experimentally investigated phase equilibria and
the phase diagrams in these materials may be classified as follows \cite
{Cahn,Hansen}:

\begin{enumerate}
\item  Systems with limited miscibility both in the liquid and solid state
(the diagram with eutectic and monotectic equilibria), for example $Pb-Zn,\
Cu-Pb,\ Ga-Pb$.

\item  Systems with limited miscibility in the solid state and unlimited
miscibility in the liquid one (the diagram with single eutectic
equilibrium), for example $Ni-Cr,\ Cu-Ag,\ Cu-CuMg_2$.
\end{enumerate}

The concentration of the second component determines the volume parts $x$
and $1-x$ of clusters having the 1 and the 2 structures respectively. Let us
study the phase diagram of the model in the~$(x,T)$~ plane (Fig.1).

The MPA equations (\ref{MPA}) may be written as
\begin{equation}
\begin{array}{c}
\tilde \omega _1=
\frac{\omega _1}x=\{1+(n-1)\exp (\frac{J_1}Tx\tilde \omega _1\frac{1-n\tilde
\omega _2}{n-1})\}^{-1} \\ \tilde{\omega }_2=\frac{\omega _2}{1-x}%
=\{1+(m-1)\exp (\frac{J_2}T(1-x)\frac{1-m\tilde \omega _2}{m-1})\}^{-1} \\
x=\{1+\frac{\tilde \omega _1}{\tilde \omega _2}\exp (\frac{J_2}T(1-x)\tilde
\omega _2-\frac{J_1}Tx\tilde \omega _1-\frac \varepsilon T(1-2x)-\frac
\alpha T)\}^{-1}.\label{MPA1}
\end{array}
\end{equation}
When ~$x=0$~ or $x=1$~, these equations coincide with those of the Potts
model for the melting of pure components. The apparent effect of the second
component is the lowering of the melting temperature. Besides that, the
redistribution of components between liquid and solid phases \cite{LANDAU}
results in two-phase domains $(L2,C2)$ and $(L1,C1)$ which are restricted by
the lines ~$x_3(T),x_4(T)$ and $x_5(T),x_6(T)$ respectively (see Fig.1). The
boundaries of a two-phase domain may be found by standard procedure from
equilibrium free energies of both coexisting phases; to find the equilibrium
free energy $F_{eq}(x,T)$, one applies the method of Lagrange multipliers:
\begin{equation}
\begin{array}{c}
-F_{eq}(x,T)=J_2(1-x)^2(
\widetilde{\omega }_2-\frac{\widetilde{\omega }_2^2}2-\frac{(1-\widetilde{%
\omega }_2)^2}{2(m-1)})+J_1x^2(\widetilde{\omega }_1-\frac{\widetilde{\omega
}_1^2}2-\frac{(1-\widetilde{\omega }_1)^2}{2(n-1)})+ \\ +\varepsilon
x(1-x)-T(1-x)\ln [\widetilde{\omega }_2(1-x)]-Tx\ln [\widetilde{\omega }_1x].%
\label{FE}
\end{array}
\end{equation}
The phase equilibrium conditions for, e.g., the domain $(L2,C2)$ are
\begin{equation}
\begin{array}{c}
F_{eq,C2}(T,x_3)-
\frac{\partial F_{eq,C2}(x_3,T)}{\partial x_3}x_3=F_{eq,L2}(T,x_4)-\frac{%
\partial F_{eq,L2}(x_4,T)}{\partial x_4}x_4; \\ \frac{\partial
F_{eq,C2}(x_3,T)}{\partial x_3}=\frac{\partial F_{eq,L2}(x_4,T)}{\partial x_4%
}.
\end{array}
\end{equation}
Here, the indexes $C2$ and $L2$ refer to the phase. For the case ~$m,n\gg 1$%
~, these conditions may be obtained in explicit form:
\begin{equation}
\begin{array}{c}
\frac{J_2}2(1-x_3^2)+\varepsilon x_3^2-\frac{J_1}{2n}x_3^2-T\ln (1-x_3)= \\
=
\frac{J_2}{2m}(1-x_4^2)+\varepsilon x_4^2-\frac{J_1}{2n}x_4^2-T\ln
(1-x_4)+T\ln m\ ; \\ T\ln
\frac{x_3}{1-x_3}+J_2(1-x_3)+2x_3\varepsilon -\frac{J_1}nx_3= \\ =T\ln \frac{%
mx_4}{1-x_4}+\frac{J_2}m(1-x_4)+2x_4\varepsilon -\frac{J_1}nx_4\label{S}
\end{array}
\end{equation}
At $x=0$ or $x=1$ these relations give (\ref{TEMP}) for the melting
temperature of pure material. For a small miscibility ($x\ll 1$)~one has
approximately
\begin{equation}
\frac{\partial x_3}{\partial T}=k_3,\ \ \ \frac{\partial x_4}{\partial T}%
=k_4,
\end{equation}
with ($m,n\gg 1$):
\begin{equation}
k_3=-\frac{2\ln m}{T_2+J_1/n-2\varepsilon },\ \ \ k_4=mk_3.
\end{equation}
The conditions determining $x_5(T)$ and $x_6(T)$ may be found in the same
way.

The miscibility gaps both in liquid and solid states and the
occurrence of two-phase domains ~$(L1,L2)$and ~$(C1,C2)$~ are due
to the ''interface
tension'' ~$\varepsilon $~ between two structures. The width of the domain~$%
(L1,L2)$~ is limited from above by the gap (fig.2). From the above equations
one obtains the relations between the parameters of the gap ~$T,x,x^{*}$~
(see fig.2) as
\begin{equation}
\begin{array}{c}
F(x)=F(x^{*}),
\nonumber \\ F(x)=-
\frac{J_2}{2m}x^2+\frac{J_1}{2n}x^2+\varepsilon x^2-T\ln (1-x)\ ;\nonumber
\\ T\ln \frac{(1-x)x^{*}}{(1-x^{*})x}=(\frac{J_2}m+\frac{J_1}n-2\varepsilon
)(x^{*}-x).
\end{array}
\end{equation}
Above the domain ~$(L1,L2)$~in the phase plane the system is homogeneous.
The line separating this domain from the homogeneous phase is a line of
phase transitions. The phase transition is of the first Ehrenfest order
anywhere except the top of the gap with coordinates $x_c,T_c$ in the $x-T$
plane
\begin{equation}
\label{critpoint}x_c=\frac 12;\ \ T_c=\frac{J_2}{4m}+\frac{J_1}{4n}-\frac
\varepsilon 2,
\end{equation}
where it is of second order. The two-phase domain $(C1,C2)$~ is limited by
lines $x_1,x_2$ (see fig.3). At low temperatures, or for large $m,n\gg 1$%
~one may approximate the equations as
\begin{equation}
\begin{array}{c}
\frac{J_2}2(1-x_1^2)-\frac{J_1}{2n}x_1^2+\varepsilon x_1^2-T\ln (1-x_1)= \\
=-
\frac{J_1}2x_2^2+\frac{J_2}{2m}(1-x_2^2)+\varepsilon x_2^2-T\ln \frac{1-x_2}%
m\ ; \\ J_2(1-x_1)-
\frac{J_1}nx_1+2\varepsilon x_1-T\ln \frac{n(1-x_1)}{x_1}= \\ =\frac{J_2}%
m(1-x_2)-J_1x_2+2\varepsilon x_2-T\ln \frac{1-x_2}{mx_2}.
\end{array}
\end{equation}

Continuing all lines obtained until their crossing points and using the
Gibbs phase rule, one gets the full phase diagram of the system. Fig.4
schematically shows the phase diagram of a system with monotectic and
eutectic equilibriums. If the crossing point of lines $x_4,~x_6$ lies in the
homogeneous phase, the phase diagram degenerates into that with single
eutectic equilibrium (see Fig.5). The values of $x_a,x_b,x^{*},T^{*}$ (see
fig.5) are related by
\begin{equation}
\begin{array}{c}
F_{eq,C2}(x_a,T^{*})-
\frac{\partial F_{eq,C2}(x_a,T^{*})}{\partial x_a}x_a= \\
=F_{eq,L}(x^{*},T^{*})-
\frac{\partial F_{eq,L}(x^{*},T^{*})}{\partial x^{*}}x^{*}= \\
=F_{eq,C1}(x_b,T^{*})-
\frac{\partial F_{eq,C1}(x_b,T^{*})}{\partial x_b}x_b; \\ \frac{\partial
F_{eq,C2}(x_a,T)}{\partial x_a}=\frac{\partial F_{eq,L}(x^{*},T^{*})}{%
\partial x^{*}}=\frac{\partial F_{eq,C1}(x_b,T^{*})}{\partial x_b},
\end{array}
\end{equation}
where the thermodynamic potentials $F_{eq}$ for corresponding phases are
defined in (\ref{FE}).

\section{Polymorphous Materials}

The model may be also used, with minor changes, to describe polymorphic
transitions in a single-component material. In this case two different local
structures are the competing local arrangements of the same atoms, so the
parameter ~$x$~ is no longer conserved. The thermodynamic potential (\ref
{fff}) and the MPA approximation (\ref{MPA1}) give the phase diagram in the $%
(\alpha ,T)$ thermodynamic plane (Fig.6). Starting from one of the phases
one changes the parameter $\alpha $ (the difference of local energies of the
two arrangements playing the role of a thermodynamic coordinate of the
state), and arrives on a polymorphic phase transition $C2\rightarrow C1$ at ~%
$\alpha \approx J_2-J_1$ in the solid state. This line may have a
continuation in the liquid area, where it terminates in the critical point:
\begin{equation}
\begin{array}{c}
\alpha _c=
\frac{J_2}m-\varepsilon +T_c\ln \frac mn-2T_c \\ T_c=\frac 14[\frac{J_1}n+%
\frac{J_2}m-2\varepsilon ]
\end{array}
\end{equation}
Below ~$T_c$~, the local orders are well distinguished, and phases ~$L1$~
and ~$L2$~ are divided by the line of the first order phase transitions with
a critical point at the end of the line (see the line AB in Fig.6). The
polymorphic liquid-liquid phase transitions were predicted and theoretically
studied earlier \cite{MP4}; recently, this prediction was confirmed by
experiments.\cite{BRAZHKIN}\cite{selen} The discovery of liquid-liquid phase
transitions in single-component materials gives a strong argument in favor
of a local order in at least some melts.

\section{Discussion of the theory}

The model proposed in this paper is rather schematic, it operates with only
a few parameters characterizing the material, and does not pretend to give a
precise quantitative description. At the same time, it gives a qualitatively
correct description of a surprisingly wide range of known phenomena, and in
particular predicts the topology of the binary system phase diagram. The
proposed model may be compared with other known models for the melting
transition.

An early model of melting is the Landau theory with a scalar order parameter
(\cite{LANDAU}, see also \cite{Kitamura}) in which a cubic term is
postulated to result in a first order phase transition. In this approach,
the nature of the transition and the short-range order in the material are
not considered. On the opposite extreme, one can calculate the thermodynamic
potentials for the liquid and crystalline phases using ''realistic''
interatomic interaction potentials\cite{Carmesin}, by numerical techniques.
Our approach is intermediate. The local order in the material has to be
found from molecular-scale calculations (or from experiment). This local
order is supposed to be a feature both of the crystalline phase and of the
melt. The possibility of crystalline local order in the melt is not expected
to be justified for all liquids; the conditions for such a feature is
discussed in \cite{PR}. Both strong covalency and local potentials of three
(or more) body type (steric interactions, hydrogen bonding, ...) favor the
local crystalline local order in the melt.. For materials with crystalline
local order both in the solid and in the molten states, the phases may be
described in terms of the local structure, using the orientation order
parameter. This also allows one to address the problem of polymorphic phase
transitions in the melt (see sec.4.2).

A special situation appears when the competing structures have substantially
different densities. The change $\alpha $ of the internal energy upon
changing the type of the local structure then includes the large work
against the external pressure. In such materials, melting may coincide with
a polymorphic transition. This may explain the ''anomalous'' melting of
covalent crystals with diamond structure ($C,Si,Ge$) and compositions of the
type ~$A_{III}B_V~(InSb,GaSb)$~ with sphalerite structure. The melting of
these substances at modest pressures is accompanied by a change of the local
order, and the coordination number has nearly 100\% increase \cite
{Glazov,Ioffe}. The large density increase causes the semiconductor - metal
transition during melting \cite{Glazov}. At high pressures, these substances
transform into metallic crystals with the structure of white ~$Sn$ \cite
{Drickamer}. One expects that there exists a line of the semiconductor -
metal transition which crosses the melting line of crystal melting. At this
point there is a three - phase equilibrium of the covalent crystal, the
metallic crystal, and the metallic liquid corresponding to the point ~$A$~
on fig.4. In principle, the coexistence line of the crystal and liquid
having different local structures may terminate (at a lower pressure) at the
crossing point of the melting line and the line of metallic liquid -
covalent liquid transition. This line of polymorphic liquid -liquid
transition extends then into the liquid area and terminates in the liquid
-liquid critical point. Recently, a liquid-liquid phase transition line of
that sort was discovered in selenium \cite{selen}. A liquid-liquid phase
transition with loss of the metallic conductivity was long ago predicted by
Landau and Zel'dovich\cite{L}.

The ''covalent'' melting takes place at high temperatures \cite{Mott}. At
lower temperatures, polymorphic phase transitions in the melt may be
expected in materials having low-coordination covalent modifications (with
the coordination number less than 4). Melting of such materials does not
require large temperatures because the energetics is determined by the
non-covalent bonds. Then, at high pressures, the covalent melt must
transform into a metal one. Recently, liquid-liquid transitions were found
in liquid $Te,\ Bi$ and $Se$ \cite{BRAZHKIN}. Qualitatively, the phase
diagram for Se has the shape shown in Fig.6. The many liquid-liquid phase
transitions in Te and Be evidence that there are several competing local
structures in the single component liquids. Compositions of the type ~$%
A_{IV}B_{VI},\ (A_V)_2(B_{VI})_3$ with covalent melting \cite{Glazov} ($%
GeTe, $ $SnTe\not ,$ $PbTe,$ $PbSe,$ $PbS,$ $Bi_2Se_3,$ $Bi_2Te_3,$ $%
Sb_2Te_3 $), may be candidates for liquid-liquid metal-covalent transitions.
If one supposes that at high pressure {\em any} substance turns out to be a
metal \cite{L}, then this phase transition should take place for the
elements of V and VI groups, but it is unclear whether it is accompanied by
a polymorphic transition. The local state representation allows one to model
the changes of the global order in condensed materials having a ''hard''
local order. This order may have components other than the structure (the
local packing of atoms, and the orientation of the local structural
anisotropy). Similar models are being used to describe magnetic \cite
{MAG,MAG1} and ferroelectric \cite{IONA} ordering. These general behaviors
can, as we have shown, be understood in terms of a very simple Potts-type
model that includes only a small number of physical parameters
characterizing the energy difference between local crystalline states, and
the energy due to interfacial interaction between two neighboring clusters
of same or different types. These parameters are physically meaningful, and
they seem to be adequate to discuss the phase diagram of binary systems.

\section{Acknowledgments}

We thank Boris I. Shumilo, Antoni C.Mitus and Michael V.Chertkov for many
discussions of crystalline ordering. This work was supported by NASA (Grant
NAG3-1932), the Chemistry Division of the ONR, and by the CAMP collaborative
MVRI program of the ONR and by the NSF-DMR through the Northwestern
University MRC (Grant DMR 9120521).

\newpage
\begin{figure}
\begin{center}
\includegraphics[width=1.0\textwidth,clip,viewport= 0 15 310 350]{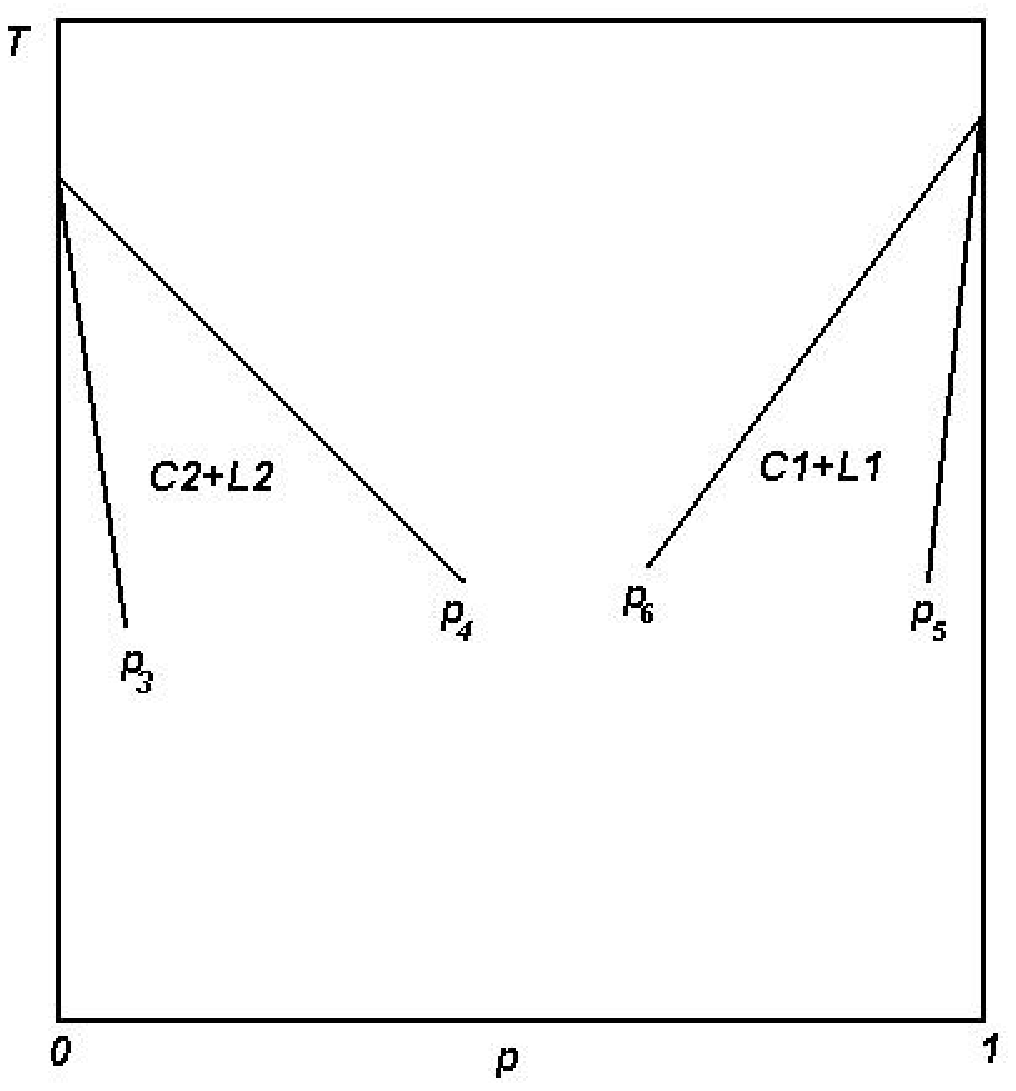}
\caption{ The thermodynamic ($x-T$) plane of a binary system. The two - phase domains
of liquid and crystal coexistence are drawn.}
\end{center}
\end{figure}

\newpage
\begin{figure}
\begin{center}
\includegraphics[width=1.0\textwidth,clip,viewport= 60 98 358 410]{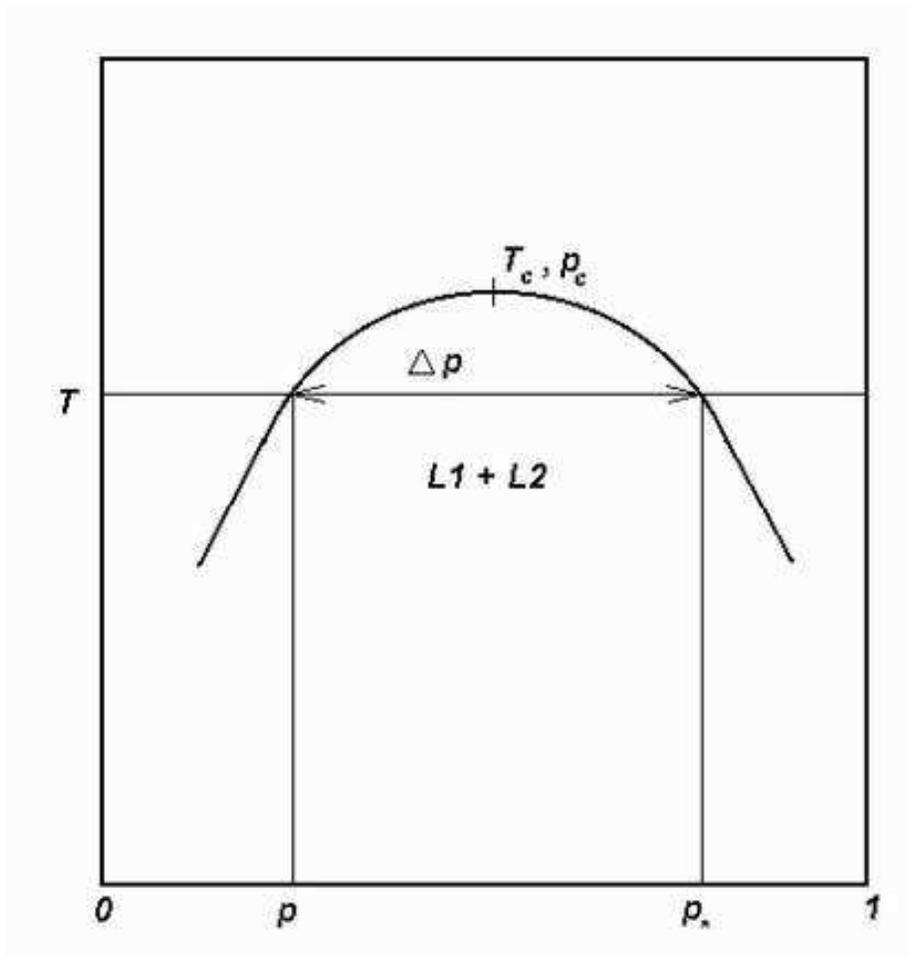}
\caption{The miscibility gap in the liquid  state. The
position of the gap corresponds to eqn. (31).}
\end{center}
\end{figure}

\newpage
\begin{figure}
\begin{center}
\includegraphics[width=1.0\textwidth,clip,viewport= 0 70 252 340]{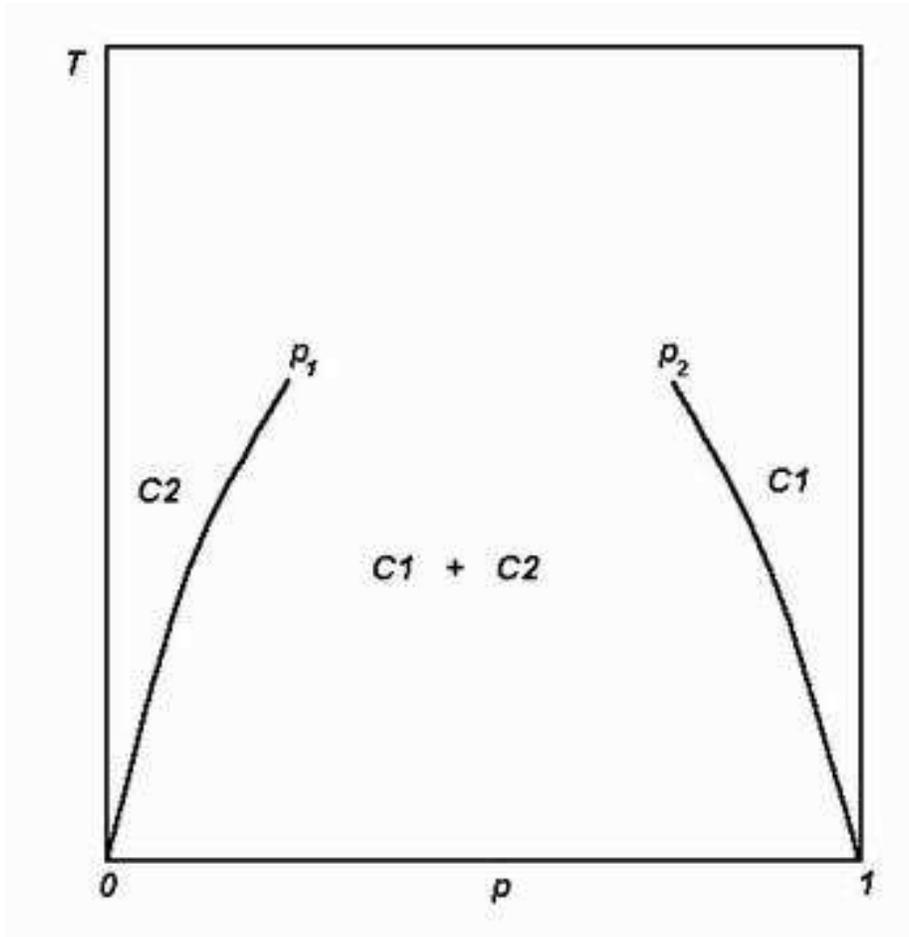}
\caption{The two-phase domain (C$_1$ and C$_2$) in the thermodynamic plane
of the system. }
\end{center}
\end{figure}

\newpage
\begin{figure}
\begin{center}
\includegraphics[width=1.0\textwidth,clip,viewport= 42 100 387 426]{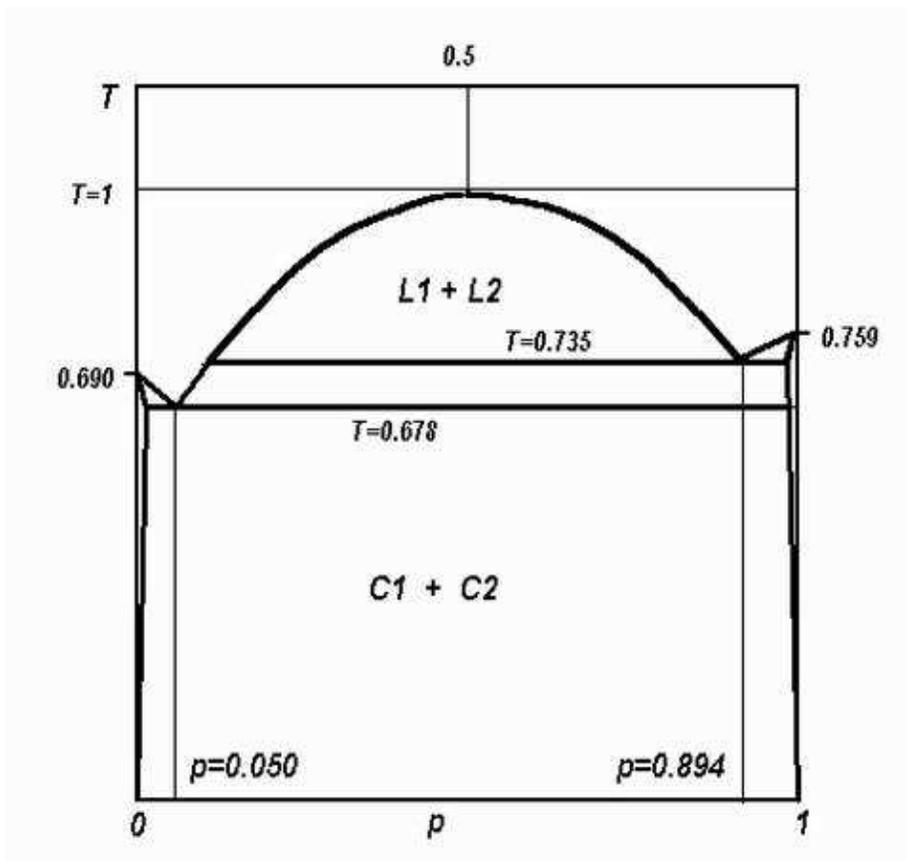}
\caption{The phase diagram of a system with monotectic and eutectic
equilibrium.}
\end{center}
\end{figure}

\newpage
\begin{figure}
\begin{center}
\includegraphics[width=1.0\textwidth,clip,viewport= 34 100 361 417]{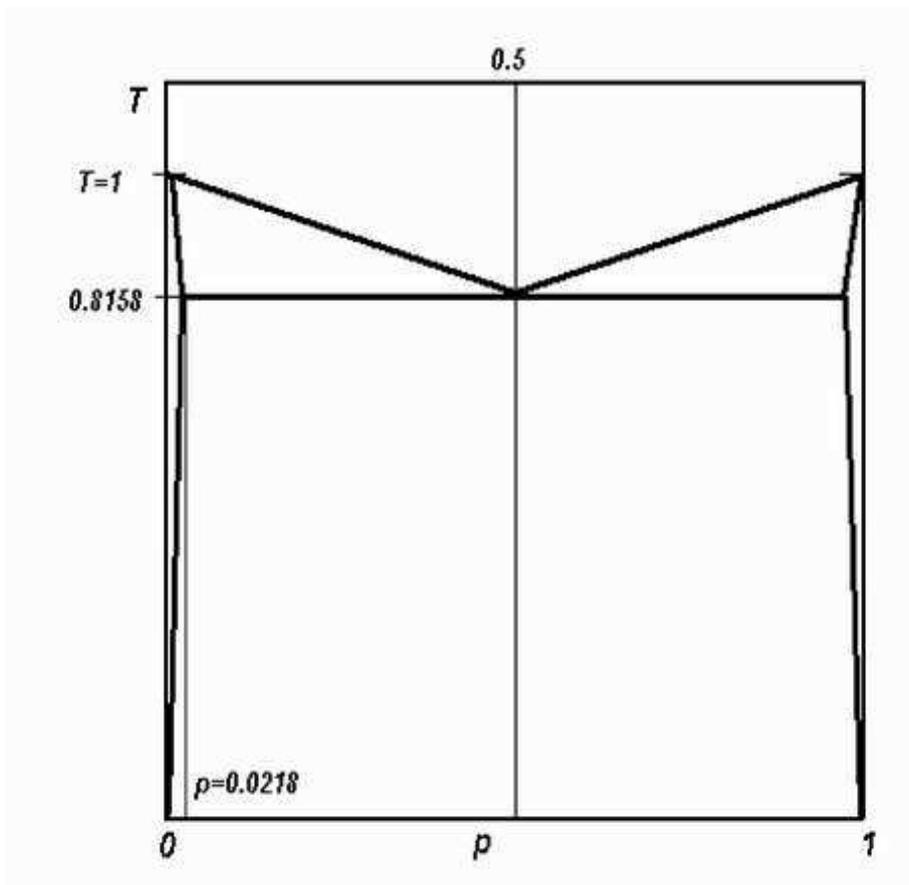}
\caption{The phase diagram of a system with a single eutectic equilibrium.}
\end{center}
\end{figure}

\newpage
\begin{figure}
\begin{center}
\includegraphics[width=1.0\textwidth,clip,viewport= 52 102 368 416]{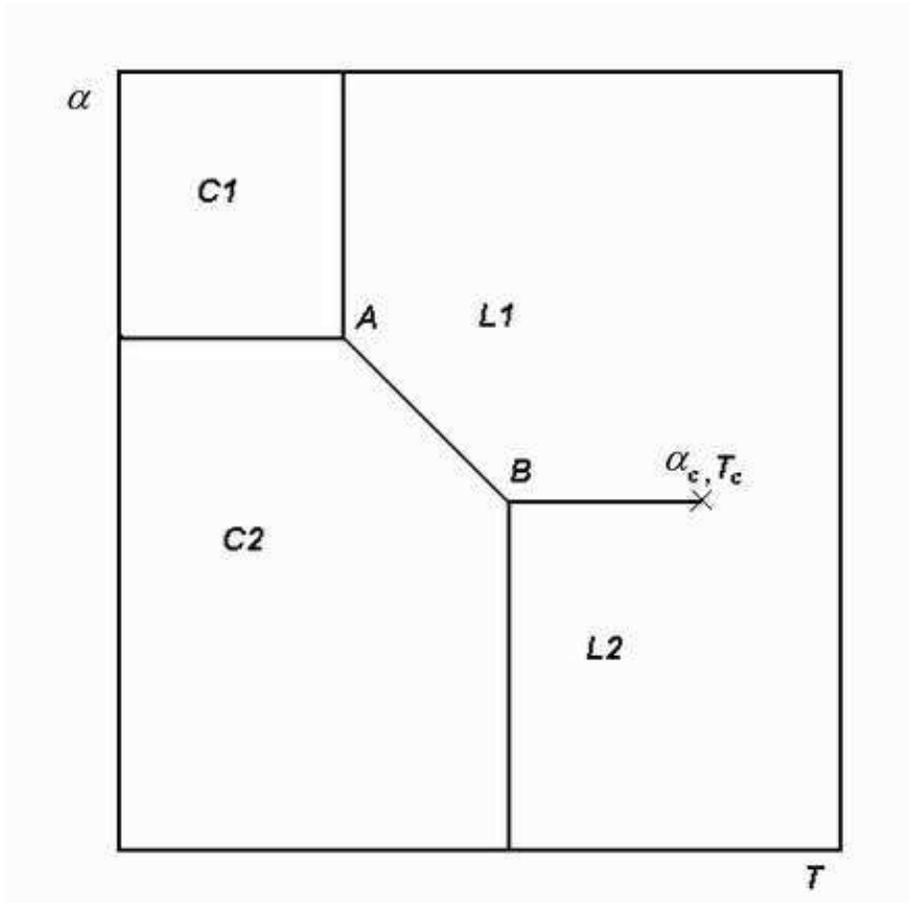}
\caption{The phase diagram of a polymorphous system.}
\end{center}
\end{figure}

\end{document}